\documentclass[twocolumn]{aa}
  \usepackage[T1]{fontenc}
  \usepackage[varg]{txfonts}
  \usepackage{graphicx}
  \usepackage{natbib}
  \usepackage{array}
  \usepackage{xspace}

\newcommand{\Nif}{$\rm ^{56}Ni$}

\def\nifsx{${}^{56}$Ni\xspace}
\def\cofsx{${}^{56}$Co\xspace}
\def\fefsx{${}^{56}$Fe\xspace}
\def\ergg{erg~g$^{-1}$\xspace}
\def\kms{km~s$^{-1}$\xspace}

\def\d{\partial}
\def\L{\left}
\def\R{\right}

\def\BE{\begin{equation}}
\def\EE{\end{equation}}
\def\BA{\begin{array}}
\def\EA{\end{array}}
\def\BAN{\begin{eqnarray}}
\def\EAN{\end{eqnarray}}

\def\FIG #1 #2 [#3] #4\par{%
 \begin{figure}
   \resizebox{\hsize}{!}{\includegraphics[#3]{#2}}
   \caption{#4}
    \label{#1}
 \end{figure}
}

\def\FIGG #1 #2 #3 [#4] #5\par{%
 \begin{figure}[!h]
   \begin{center}
   \includegraphics*[#4]{#2}
   \includegraphics*[#4]{#3}
   \caption{\label{#1}#5}
   \end{center}
 \end{figure}
}

\def\FIGs #1 #2 #3 #4 #5 [#6] #7\par{%
 \begin{figure}[!h]
   \begin{center}
       \includegraphics*[#6]{#2}
       \includegraphics*[#6]{#3}
       \includegraphics*[#6]{#4}
       \includegraphics*[#6]{#5}
       \caption{\label{#1}#7}
   \end{center}
 \end{figure}
}

\def\FIGss #1 #2 #3 #4 #5 #6 #7 [#8] #9\par{%
 \begin{figure}[!h]
   \begin{center}
       \includegraphics*[#8]{#2}
       \includegraphics*[#8]{#3}
       \includegraphics*[#8]{#4}
       \includegraphics*[#8]{#5}
       \includegraphics*[#8]{#6}
       \includegraphics*[#8]{#7}
       \caption{\label{#1}#9}
   \end{center}
 \end{figure}
}

\begin{document}

\title{Theoretical light curves for deflagration models of Type~Ia supernova} 
\titlerunning{Synthetic light curves of SNe~Ia}

\author{S.I.~Blinnikov\inst{1,2}, F. K. R{\"o}pke\inst{1}, E.I.~Sorokina\inst{1,3},         M.~Gieseler\inst{1}, M.~Reinecke\inst{1}, C.~Travaglio\inst{4} ,
        W.~Hillebrandt\inst{1}  \and M.~Stritzinger   \inst{5,1} }                     
 \authorrunning{S.I.~Blinnikov et al.}         
   \offprints{S.I.~Blinnikov}

   \institute{Max-Planck-Institut f\"ur Astrophysik,
              Karl-Schwarzschild-Str. 1, D-85741 Garching, Germany\\
             \email{[seb;fritz;elena;mccg;martin;wfh;stritzin]@mpa-garching.mpg.de}
         \and
             ITEP, 117218 Moscow, Russia \\
           \email{sergei.blinnikov@itep.ru}
         \and
             Sternberg Astronomical Institute, 119992 Moscow, Russia \\
                  \email{sorokina@sai.msu.su}
         \and
              INAF -- Osservatorio Astronomico di Torino,
              Strada dell'Osservatorio 20, I-10025 Pino Torinese,
              Torino, Italy\\
              \email{travaglio@to.astro.it}
         \and{Dark Cosmology Centre, Niels Bohr
   Institute, University of Copenhagen, Juliane Maries Vej 30, DK-2100
   Copenhagen \O, Denmark}
             }
   \date{Received November ..., 2005; accepted ...}

  \abstract
  {} 
   {We present synthetic bolometric and broad-band $UBVRI$
light curves of SNe~Ia, 
for four selected 3-D deflagration models of 
thermonuclear supernovae. }
   {The light curves are computed with the 1-D hydro code {\sc
  stella}, which models
(multi-group time-dependent) non-equilibrium radiative transfer inside 
SN ejecta. Angle-averaged results from 3-D hydrodynamical explosion
simulations with the composition determined in a nucleosynthetic
postprocessing step served as the input to the radiative transfer
  model.}
   {The predicted model $UBV$ light curves do agree
reasonably well with the observed ones for SNe~Ia in the range of low
to normal luminosities, although the underlying hydrodynamical
explosion models produced only a modest 
amount of radioactive \nifsx (i.e. $\sim$ 0.24 - 0.42\,M$_{\sun}$) 
and relatively low kinetic energy in the explosion (less than $0.7
\times 10^{51}$ erg).
The evolution of predicted $B$ and $V$ fluxes 
in the model with a \nifsx mass of 0.42 M$_{\sun}$ follows 
the observed decline rate after the maximum very well,  
although the behavior of fluxes in other filters somewhat deviates
  from
observations, and the bolometric decline rate is a bit slow.
The material velocity at the photospheric level is of the order of
  $10^4$
km~s$^{-1}$ for all models.
Using our models, we check the validity of Arnett's rule, relating the
  peak
luminosity to the power of the deposited radioactive heating, and we
  also check
the accuracy of the procedure for extracting the \nifsx mass
from the observed light curves.}
   {We find that the comparison between theoretical light curves and
observations provides a useful tool to validate SN~Ia models.
The steps necessary to improve the agreement between
theory and observations are set out.}

 
\maketitle

\section{Introduction}
\label{intro}

With three-dimensional simulations 
\citep[e.g.][]{hillebrandt2000b,reinecke2002d,gamezo2003a,roepke2005b}, 
type~Ia supernova (hereafter SN~Ia) explosions can be modeled
self-consistently. Such models are constructed from
first principles, avoiding free parameters in the description of physical
processes. The only parameters entering are
the initial state of the exploding white dwarf star and the
configuration of flame ignition, which must be determined in
separate studies of progenitor evolution.

Consequently, the important question is to what degree do these models
describe real SNe~Ia explosions. This can only be answered through comparison
with observations of nearby events 
(e.g. \citealt{pignata04,benetti04,stehle05}). 
To this end, observables need to be derived from theoretical
explosion models. In this paper, we show how synthetic light curves
(both bolometric and broad-band UBVRI) can be obtained from a set of three-dimensional
deflagration models of SN Ia explosions (\citealt{Trav2004,roepke2005,roepke2005b}). 
We emphasize that the present study is
intended to demonstrate the derivation of light curves from
theoretical models and to analyze the use of these as a tool to
validate such models. Therefore the chosen models do not reflect
particularly realistic simulations and have known weaknesses.
Judging the validity of explosion models in terms of  
derived observables requires more elaborate models and this issue will
be addressed in a forthcoming publication.

When one succeeds in constructing SN~Ia explosion models that are 
consistent with
observations, they will be used to tackle questions concerning
the application of SNe Ia in determining cosmological
parameters. Since these objects need to be calibrated on the basis of
empirical recipes to serve as distance indicators, such distance measurements
are subject to uncertainty. This may be lessened by a
theoretical understanding of the origin of the diversity in
observables and of the calibration techniques.

The calibration methods currently applied in SN~Ia cosmology are based on
an empirical correlation between the blue-band magnitude of the
supernova and the decline rate of its light curve
\citep[e.g.][]{pskovskii77,phillips93,phillips99}. 
As pointed out by many authors \citep[e.g.][]{hwt98,Sorokina00,Mannucci},
the luminosity--decline rate relation for SNe~Ia  is derived 
at low redshifts $z$, and it is unclear how it should change at 
high redshifts. 
Systematic differences of the order of
$\sim$$0.1$ magnitudes \citep{hoflich00,nomoto03} 
can be important in cosmological applications.
A sound understanding of the
physical mechanism of SNe~Ia is crucial in any project investigating
possible systematic trends in their properties.
Additionally, a radiative transfer algorithm is required to reliably relate the 
hydrodynamical model of the explosion to observed fluxes and spectra of
SNe~Ia.

\citet{roepke2004c} and \citet{roepke2005}
have presented systematic studies of effects resulting from variations
in the progenitor white dwarf's carbon-to-oxygen ratio, its central
density at ignition and its metallicity on 3-D
deflagration explosion models. These were based on simple and thus
only weakly exploding models and therefore such studies need to be
extended with more elaborate simulations. The missing link to SN~Ia
cosmology in such approaches is the derivation of synthetic light curves. 

This paper presents synthetic light curves for two models from the set
described by \citet{roepke2005} and for two additional 3-D models
\citep[from][]{Trav2004, roepke2005c}.
We compute bolometric and broad-band $UBVRI$
light curves of SNe~Ia, employing the radiation hydro code 
{\sc stella} which simulates multi-group time-dependent non-equilibrium
radiative transfer inside SN ejecta in a spherically-symmetric
approximation. 

First, we briefly outline the assumptions made in modeling
the radiative transfer (Sec.~\ref{radmodel}).  
In Sec.~\ref{param} we describe parameters of the hydrodynamical models.
Various observables predicted by the radiation code
for the studied hydro models are presented in Sec.~\ref{observ}.
The accuracy of determining the $^{56}$Ni mass
based on Arnett's rule is demonstrated in Sec.~\ref{ni56}.

The results are summarized and discussed in Sec.~\ref{discuss}.
We describe the necessary steps for future improvements
of both hydrodynamical and radiative transfer models
in Sec. 7.

\section{Numerical Model for Radiative Transfer}
\label{radmodel}

Modeling the post-explosion stages of SNe~Ia
appears to be easier than that 
of core collapse supernovae (CCSNe). For thermonuclear supernovae 
the hydrodynamics is very simple:
the coasting stage starts very early, there are no shocks, and the only 
additional heating mechanism is the decay of radioactive nuclei.
A typical  assumption for SNe~Ia light curve modeling is to fully
neglect the hydrodynamics.
We discuss the validity of this assumption in Sect.~\ref{discuss}.

However, with SNe~Ia difficulties arise with the modeling of 
the radiative transfer.
SNe~Ia become almost transparent in the continuum at the age of a few weeks.
This means that NLTE effects are stronger compared to CCSNe.
Radiation decouples from 
matter within the entire SN~Ia ejecta prior to maximum light
(\citealt{EThn,blsorprt}).
In this environment one cannot ascribe the gas temperature (nor any other
temperature) to radiation since a SN~Ia spectrum differs strongly from a 
blackbody.
Instead, one has to solve a system of time-dependent transfer equations
in many energy groups
with an accurate prescription for the treatment of a enormous number 
of spectral lines. These lines are the main source of opacity in the 
ejecta (\citealt{BHM,PE00b}).

Recently, powerful codes have been developed  in order to address
the full 3-D time-dependent problem of SN~Ia radiative transfer 
(\citealt{hoeflich02,Lucy05}).
Yet there are some basic questions, like averaging the line opacity
in an expanding media, that remain controversial.

We compute broad-band $UBVRI$ and bolometric light curves of SNe~Ia with the
multi-energy radiation hydro code {\sc stella} (\citealt{blinn98,blsoruv}).
The 3-D hydrodynamic models are angle-averaged
and used as input for the radiation hydro code.
Time-dependent equations for the angular moments of intensity in fixed
frequency bins are coupled to the Lagrangian hydro equations and solved
implicitly.
Therefore there is no need to ascribe any temperature to the radiation;
the photon energy distribution may be quite arbitrary.

There are millions of spectral lines that lead to the formation 
of a SN~Ia spectrum, and it is
not a trivial problem to find a convenient way to treat them
even in the static case.
The expansion makes the problem much more difficult to solve as 
hundreds to even thousands of lines give their input into emission and
absorption at each frequency.

The effect of line opacity is treated in the current work as an expansion opacity according to
\citet{EP93}. The line list is limited to $\sim$160 thousand entries, selected
from the strongest down to the weakest lines until the saturation in the
expression for expansion flux opacity is achieved.
The effect of extending the line list deserves further investigation, and
work in this direction is underway.

The ionization and atomic level populations are described by Saha-Boltzmann expressions.  However, the source function is {\em
not\/} in complete LTE.

The source function at wavelength $\lambda$ is
\BE
S_\lambda = \epsilon_{\rm th}\,B_\lambda + (1-\epsilon_{\rm th})\,J_\lambda,
\label{S}
\EE
where $\epsilon_{\rm th}$ is the thermalization parameter, $B_\lambda$ 
is the Planck function, and $J_\lambda$ 
is the angle mean intensity.
In a pure LTE approximation $\epsilon_{\rm th}=1$, and in
a ``pure scattering'' treatment $\epsilon_{\rm th}=0$. 
\citet{BHNB}  compared 
the results of a full NLTE treatment, a pure LTE treatment, and a pure scattering treatment of the lines (in their case $S_\lambda$ described the source function
only in a line). They found that the pure LTE case (for lines) reproduces
the overall spectral shape of the NLTE case rather well, while the
pure scattering case does not.
In the NLTE case, the collisions within multiplets create a
pseudo-thermal pool of photons which are in equilibrium among themselves.

In our code, we have rather wide energy bins that contain hundreds of strong
lines, so our $S{_\lambda}$ is a complicated average of lines and continuum for
a given bin.
We can ascribe an arbitrary degree of thermalization to the lines.
Even if we treat all lines as pure absorbers we have an appreciable
contribution of photon scattering off electrons, thus the 
average $\epsilon_{\rm th}\ne 1$ in a bin and $S_\lambda$ is not in full LTE.
Therefore, we prefer to call the approximation of full thermalization
in lines as `fully absorptive lines'.
The approximation of the fully absorptive opacity in spectral lines allows
us to simulate NLTE effects in a simple manner.
NLTE results (\citealt{BHNB}) and the ETLA approach (\citealt{PE00b}) 
demonstrate that fully absorptive lines give
an acceptable description of the overall spectral shape and subsequent
optical light curves.

A similar approach to light curve modeling with LTE opacities and a
comparable number of photon energy groups is used by \citet{hwt98}
who employed NLTE calculations \citep{hoeflich95} to calibrate scattering in lines.
A full NLTE time-dependent treatment is within reach and is already being implemented
by other groups (\citealt{hoeflich02,Lucy05}), but  
much work remains to be done before all problems are solved.

The deposition of gamma rays produced in radioactive decays of Fe-group
isotopes (mostly due to $^{56}$Ni and $^{56}$Co) should be properly considered.
After being emitted, the gamma rays travel through the ejecta and can
end up either thermalized or in non-coherent scattering process.
To determine which way they end up one has to solve
the transfer equation for gamma rays together with hydrodynamical
equations.
The full system of equations should also involve the radiative transfer 
equations in the range from soft X rays to infrared wavelengths
for the expanding medium.

We do not write down the full set of equations solved by {\sc stella} 
(\citealt{blinn98}), but we point out that hydrodynamics coupled to radiation
is fully computed (homologous expansion is not assumed).
Since there are several different approaches found in literature, here 
(and in Sect.~\ref{discuss}) we discuss in some detail the use of the 
matter temperature equation.

The equation for the material temperature $T$ is
derived from the first law of thermodynamics,
\BE
de+pdV = de-{p\over\rho^2}\,d\rho=
                        \L({\d e\over\d T}\R)_\rho\,dT
                 -{T\over\rho^2}\,\L({\d p\over\d T}\R)_\rho\,d\rho,
\label{de}
\EE
which takes the form:
\begin{equation}
 \biggl({\partial e\over \partial T}\biggr)_\rho{\partial T\over \partial
 t} =\varepsilon+4\pi \int\limits_0^\infty
 (\alpha_{\nu}J_\nu-\eta_\nu)d\nu 
  - 4\pi {\partial{(r^2u)}\over \partial m} T
 \biggl({\partial p\over \partial  T}\biggr)_\rho \;.
\label{tempr}
\end{equation}
Here $p=p(\rho,T)$ is the material pressure, $\rho$ is the density, $m$
is the Lagrangian coordinate (i.e. the mass inside radius $r$),
$J_\nu$ is equal to ${1\over 2}\int_{-1}^1 d\mu\,I_\nu $,
 $e$ is the specific internal energy of matter, $\alpha_{\nu}$ is the
absorption coefficient per unit mass, and $\varepsilon$ is 
the specific power of the local heating (here due to gamma-ray 
energy deposition). In the case of LTE the emission coefficient $\eta_\nu$
is equal to $\alpha_{\nu}B_\nu$. 

We should note that the last equality in Eq.~(\ref{de}) is based on the
expression $de+pdV = Tds$, where $s$ is specific entropy, i.e. not
only the first, but also the second law of thermodynamics for
reversible processes
is used in the derivation.
In other words, this relation is valid when $e$ and $p$ are functions of $\rho$ and $T$.
While radiation is non-equilibrium in our approximation, ionization and
atomic level populations are described by Saha-Boltzmann equations. Therefore
Eqs.~(\ref{de}) and (\ref{tempr}) are applicable. 
If ionization and excitation depend on other parameters in NLTE,
and/or are described by kinetic equations, one must use other forms
of the energy equation (see e.g. \citealt{Sorokina04}). 

Another possible approach is to extract matter energy $e$ 
from the equations of evolution for radiation energy $E$ and full energy $e+E$
(e.g. \citealt{hoeflich93}). 
In both cases the matter energy $e$ is found with lower accuracy than
the radiation energy $E$. This is due to the subtraction of nearly equal terms; 
either of $e+E$ and $E$ in the latter approach or in the heating and cooling
terms of Eq.~(\ref{tempr}).

To find the radioactive energy deposition $\varepsilon$ 
we treat the gamma-ray opacity as a pure absorptive one, and solve the
gamma-ray transfer equation in a one-group approximation 
following~\cite{SSH}.
The effective opacity assumed is 
$\kappa_\gamma=0.05 Y_e \; \mbox{cm}^2/\mbox{g}$, where $Y_e$ is the total electron number density over baryon density.
We have checked that this effective opacity gives good agreement for the values 
of the deposition and broad band fluxes found by Monte-Carlo codes after the maximum light. The rise part of the light curve is not sensitive to the variation
of $\kappa_\gamma$ when it is changed by a factor of 3.
For one of the models (\emph{b30}, described below) the deposition was computed in full 3D-transport by a modification of {\sc shdom} code \citep{evans}.
The agreement with 1D version is within a few percent.

The heating by the decays  $^{56}$Ni  $\to$ $^{56}$Co$\to$ $^{56}$Fe is
taken into account. It is assumed that positrons, born in the decays, are
trapped and deposit their kinetic energy locally, cf.  
\citet{Colgate80a,Colgate80b,ChanLing93,Pilar97,Milne97,PilarHenk98}
and \citet{Jeffery99} on various approaches to this problem
important for late stages (the `tail') of SNe~Ia. 

To calculate SNe~Ia light curves our method can use up to 200
frequency bins and up to $\sim$400 zones in mass as a Lagrangian coordinate on
a modest processor. 

Major recent improvements in the code {\sc stella} \citep[introduced after the paper][was published]{blinn98}
are described in the following.
The ionization of the 15 most abundant elements (from H up to Ni) is now computed  
for any stage, without introducing an averaged ion approximation.
As previously mentioned the line list has been extended. When computing the opacity
we now take into account the change in the composition of the iron peak elements
in the decay chain $^{56}$Ni  $\to$ $^{56}$Co$\to$ $^{56}$Fe. 
The most important advancement is also related to the computation of the expansion
opacity: instead of interpolating the opacity in tables for preselected
instants of time, it is now computed for each time step in each mesh zone.
Although this approach requires much more processor time than using the
tables, it is more reliable when one is interested in finding fine details
for models with small differences in parameters.

\section{Initial models}
\label{param}

For this study light curves are derived using four hydrodynamical
explosion models. Each model possesses different characteristics that
result in a variation in the computed light curves.

The explosion models are based on a pure deflagration scenario. 
In this case a sub-sonic thermonuclear
(``deflagration'') flame is ignited near the center of 
a Chandrasekhar mass carbon-oxygen white dwarf. Subsequently the flame
propagates outwards generating turbulence due to generic 
instabilities.
The interaction with
turbulent motions accelerates the flame propagation velocity such that
nuclear burning releases a sufficient amount of energy that 
completely disrupts the white dwarf. For a review on SN~Ia models see
\citet{hillebrandt2000a}.

Details concerning the numerical techniques used to implement this scenario 
were presented by \citet{niemeyer1995b}, \citet{reinecke1999a,reinecke2002b} 
and \citet{roepke2005c}. The main challenge is
to self-consistently model the turbulent combustion which involves a
wide range of scales. To this end a Large Eddy Simulation
(LES) approach is applied, which directly resolves only the largest
scales of the problem (i.e. scales from the radius of the white dwarf,
$\sim$2000 km, down to several kilometers). To 
account for the effects of turbulence on unresolved scales, a
sub-grid scale model is adopted (\citealt{niemeyer1995b}). 
The flame itself is modeled as a sharp
discontinuity separating the fuel from the ashes applying the
level-set method \citep{reinecke1999a}.
Since such an approach does not resolve the internal
flame structure, propagation due to burning has to be prescribed. This approach
however does not lead to any free parameters. The theory of turbulent
combustion states that in the case of burning in the so-called
flamelet regime, which applies to most parts of a SN~Ia explosion,
the flame propagation completely decouples from the micro-physics. It
is solely determined by the turbulent motions which can be derived
from the adopted sub-grid scale model.  
We emphasize that apart from the initial
conditions that are determined from the progenitor evolution, this model
contains no free parameters. 

Consequently, the models used to derive the light curves from vary only in
the initial conditions, i.e. the composition, the central density of
the white dwarf, and the manner in which the thermonuclear flame is
ignited. All models were simulated on one spatial octant of the
white dwarf and mirror symmetry was assumed for the other
octants. \citet{roepke2005b} showed that this artificial symmetry
constraint does not obscure the explosion
mechanism\footnote{Asymmetric explosions with possible implications on
  light curve modeling may, however, develop from asymmetries in the flame
  ignition configuration.}. In three of the
models, the flame was ignited centrally and perturbed from spherical
symmetry by toroids. Contrary to this, the  \emph{b30\_3d\_768} model of 
\citet{Trav2004} (in the following denoted as \emph{b30}), assumed
a multi-spot ignition scenario in which the flame was ignited in 30 small
kernels per octant. Evidently, such multi-spot scenarios can lead to more
vigorous explosions than centrally ignited models
(\citealt{Trav2004,roepke2005d}). 

\begin{table}
\centering
\caption{Model parameters.
\label{models_tab}}
\setlength{\extrarowheight}{2pt}
\begin{tabular}{llllll}
\hline\hline
model & $\rho_{c9}$ & $X(^{16}\textrm{O}) $ & 
$X(^{12}\textrm{C}) $ & $X(^{22}\textrm{Ne})$ & $Z[Z_{\sun}]$  \\
\hline
\emph{1\_3\_3}  & $1.0$ & $0.38$  & $0.545$  & $0.075$  & $3.0$ \\
\emph{2\_2\_2}  & $2.6$ & $0.54$  & $0.435$  & $0.025$  & $1.0$ \\
\emph{c3\_3d}   & $2.9$ & $0.5$   & $0.475$  & $0.025$  & $1.0$\\
\emph{b30}      & $2.9$ & $0.5$   & $0.475$  & $0.025$  & $1.0$ \\
\hline
\end{tabular}
\end{table}

The initial parameters of the models are summarized in
Table~\ref{models_tab}. We assumed 
increasing central densities in the sequence of models \emph{1\_3\_3},
\emph{2\_2\_2} and \emph{c3\_3d}. Explosion models
\emph{c3\_3d} and \emph{b30} were set up for a 50\% carbon-oxygen
mixture, while the carbon mass in models \emph{2\_2\_2} and
\emph{1\_3\_3} was set to 0.46 and 0.62, respectively.

The detailed nucleosynthesis of each model was derived
with a post-processing technique that uses data obtained from tracer
particles in the explosion simulation 
\citep[see][ for a detailed discussion concerning
this method]{Trav2004,roepke2005}.
During the post-processing solar
metallicity was assumed for all the models except \emph{1\_3\_3}. There an assumed 
three times increase metallicity of the main sequence progenitor of
the white dwarf resulted in a $^{22}$Ne mass fraction of 0.075. 
The $^{22}$Ne mass fraction was included in the post-processing by lowering 
the $^{12}$C fraction.

With respect to the computational setup, model \emph{c3\_3d}
differs from the other three models. These were calculated on a static grid
with uniform fine-resolved inner parts and exponential grid spacing
in the outer regions to capture at least parts of the expansion. Models
\emph{b30}, \emph{2\_2\_2}, and \emph{1\_3\_3} were followed for
$1.33 \, \mathrm{s}$, $1.5 \, \mathrm{s}$, and $2.0 \,
\mathrm{s}$, respectively. The light curve derivation assumes
homologous expansion at the point where the hydrodynamical explosion
simulation ends. This is, however, not yet reached to a high accuracy at these
times. Therefore, in model \emph{c3\_3d} a different approach was
chosen (\citealt{roepke2005c}). This simulation was carried out on a uniform
computational grid co-expanding with the white dwarf. In this way it was
possible to follow the evolution time to 10 s after flame ignition.

All simulations (except \emph{b30}) 
were carried out on $[256]^3$ grid cells to 
ensuring numerical convergence (\citealt{reinecke2002b,roepke2005c}). 
To accommodate for a reasonable number of flame ignition kernels the 
resolution was increased to $[756]^3$ cells for model \emph{b30}.

The global characteristics of the explosion models are summarized in
Table~\ref{energ_tab}. Detailed results from the explosion simulations
can be found in 
\citet{Trav2004}, \citet{roepke2004c,roepke2005b} and \citet{roepke2005}\footnote{\citet{kozma2005a}
derived a nebular spectrum for model \emph{c3\_3d}. In their case this model was 
denoted \emph{c3\_3d\_256\_10s}.}
All models produce rather weak explosions, lower than canonical $1$ foe ($=10^{51}$ erg), and
low amounts of \nifsx. Although they lie in the range of variability
expected from observations, only model \emph{b30} comes close to the
class of ``Branch normals'' (\citealt{branch1993a}).

The remapping of 3-D models to spherical 1-D geometries has been done using
the same tracer particles that were used for the nucleosynthesis
calculations.
A grid in radial coordinates was constructed with uniform steps in radius,
the outer radius being equal to the maximum radius of the 3-D model.
The number of tracer particles (all of equal mass) then determined
the mass of each spherical shell constructed on the grid, and
the chemical composition of the particles determines the composition of
the shell.
The square of the particle velocities were summed up to obtain the total
kinetic energy of the shell.
The motion of particles is not purely radial, however, the non-radial component
is not high at the end of explosion simulation. The kinetic energy of 
transversal motion is of the order of $0.01$ of the radial motion 
(since model \emph{c3\_3d} was followed for a longer period the 
transversal motion is only $\sim 3\times 10^{-4}$ of the radial motion).

We assume that even if the non-radial component will be dissipated and goes
to heat, the heat will not be radiated away because during the first several
hours after the explosion all zones are optically thick. 
After some time this heat will be returned to the kinetic energy of the 
expansion.
The kinetic energy and the mass of each shell determines the velocity
of the shell, which is ascribed to the outer radius of the mesh zone
in the radiation hydro code. 
The radiation hydro simulation starts for all runs at time $10^{4}$ s
after the ignition of the explosion.
We assume that the total energy at the end of the flame simulation
$E_{\rm tot}$ (kinetic plus thermal plus gravitational) goes into
$E_{\rm kin}$ at $10^4 \, \mathrm{s}$. 
We have renormalized the original $E_{\rm kin}$ 
from the tracer particles to  $E_{\rm tot}$, and this value is given
in the Table~\ref{energ_tab} as $E_{\rm kin}$ initial. 

As an example, the chemical composition for the model \emph{2\_2\_2}
is shown in Fig.~\ref{chemm222}. Here the mass fractions of
the most abundant elements are presented 
as a function of the mass inside each shell.
This distribution of the
composition is preserved during the radiation hydro run for all elements 
(except for \nifsx which decays successively to \cofsx and \fefsx).
A better presentation of the innermost and outermost layers is achieved
when the composition is plotted as a function of the material velocity.
This is shown for all models in Figs.~\ref{chem133}-\ref{chemb30}. 
However, since not all of the initial models are in full homology, 
the velocity is not a good  Lagrangian coordinate. 
One can notice  in Fig.~\ref{chem133} that some outer shells
have velocities lower than the adjacent inner shells.
Moreover,
the homology is never perfect during first weeks due to the heating
from the \nifsx $\rightarrow$ \cofsx $\rightarrow$ \fefsx decay chain.
Nevertheless, the plots in  Figs.~\ref{chem133}-\ref{chemb30} do show
the expected model distribution of chemical elements at the observable stages
of supernova evolution with an accuracy of $\sim 1$\%.

Due to noise in the composition and density when averaging the 3-D model,
the number of radial zones was restricted to 50
(see, e.g. Fig.~\ref{chemm222} and  Fig.~\ref{rhoc3}).
No smoothing of the composition or the density in the radial direction has been
done during the 3-D to 1-D remapping. Note the value of any quantity  in a 
spherical shell is just a mean value over $4\pi$ for the same shell in the 
3-D model.

\begin{table}
\centering
\caption{Mass of $^{56}\textrm{Ni}$ in the models
and their energetics
\label{energ_tab}}
\setlength{\extrarowheight}{2pt}
\begin{tabular}{llll}
\hline\hline
model & $M(^{56}\textrm{Ni}) [M_{\sun}]$ & $E_{\rm kin}$, foe, initial & $E_{\rm kin}$, foe, asymptotic \\
\hline
\emph{1\_3\_3}   & $0.24$ & 0.357  &  0.365 \\
\emph{2\_2\_2}   & $0.31$ & 0.441  &  0.453 \\
\emph{c3\_3d}    & $0.28$ & 0.431  &  0.441 \\
\emph{b30}       & $0.42$ & 0.663  &  0.679 \\
\hline
\end{tabular}
\end{table}

Table~\ref{energ_tab} presents the parameters that are the most important
for light curve modeling. These include: (1) the \nifsx mass and
(2) the initial and final kinetic energy.
Note that the asymptotic kinetic energy is somewhat higher than 
the initial kinetic energy
due to the addition of energy from the radioactive decay of \nifsx. 
This small effect is discussed below (see Sect.~\ref{hydroia}). 

\FIG chemm222 chemmpa222 []
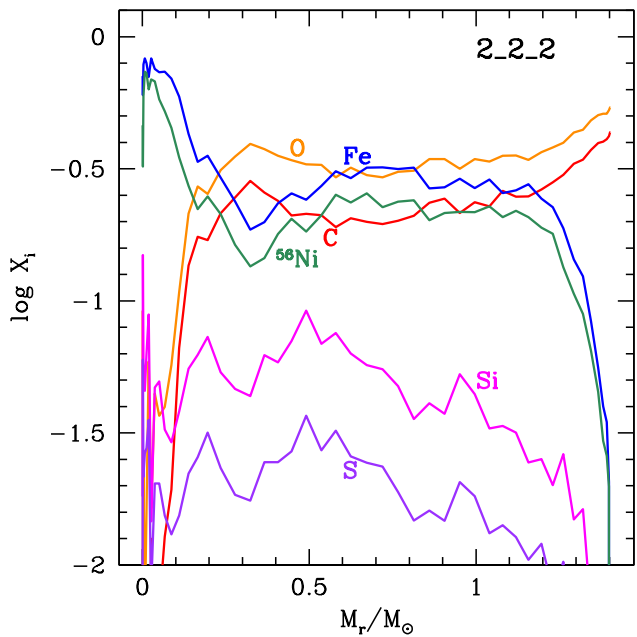
Mass fractions of the most abundant elements produced in model \emph{2\_2\_2} 
as a function of Lagrangian mass. Note `Fe' includes iron-peak elements 
together with $^{56}$Ni. 

\FIG chem133 chemv133 []
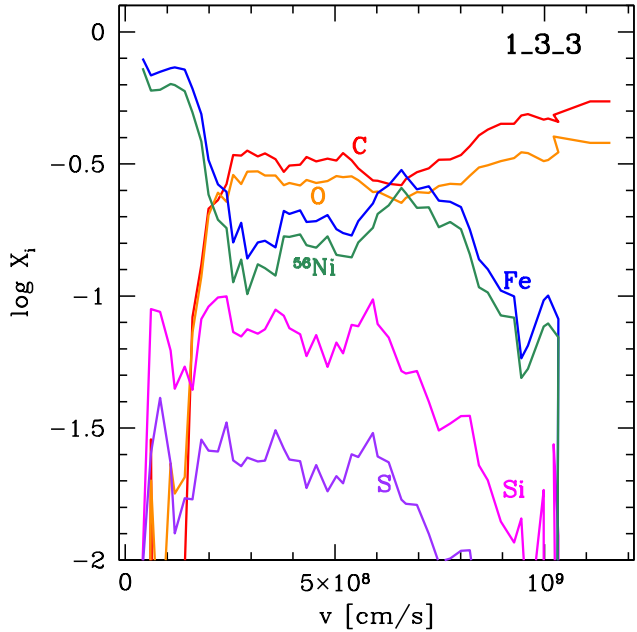
Mass fractions of most abundant elements for the model \emph{1\_3\_3}
as a function of the material velocity.

\FIG chemv222 chemv222 []
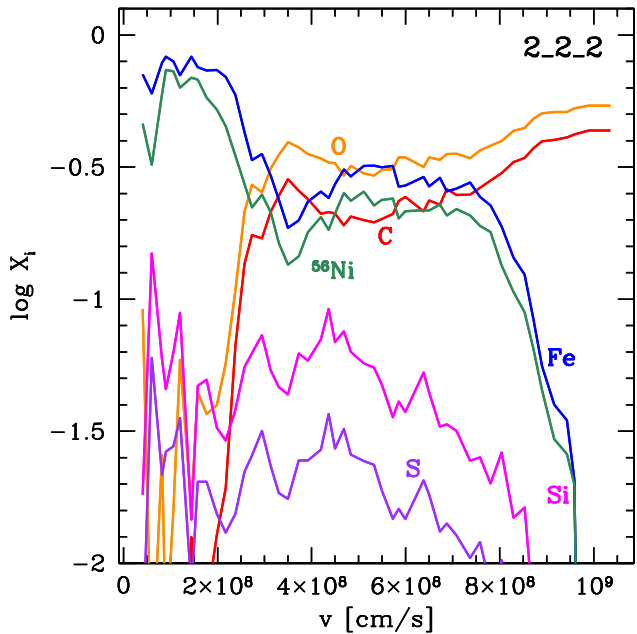
Mass fractions of the most abundant elements produced in model \emph{2\_2\_2}  
as a function of material velocity. Note `Fe' includes iron-peak elements 
together with $^{56}$Ni.  

\FIG chemc3 chemvc3 []
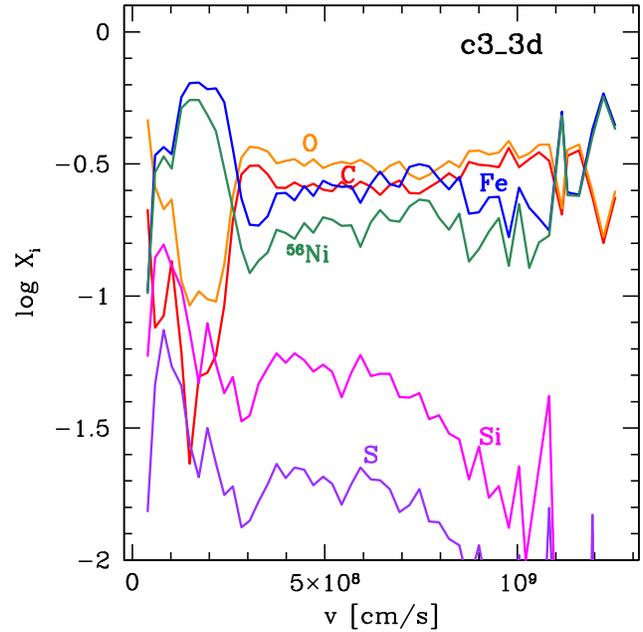
Mass fractions of most abundant elements for the model \emph{c3\_3d}
as a function of the material velocity. Note `Fe' includes iron-peak elements
together with $^{56}$Ni.

\FIG chemb30 chemvb30 []
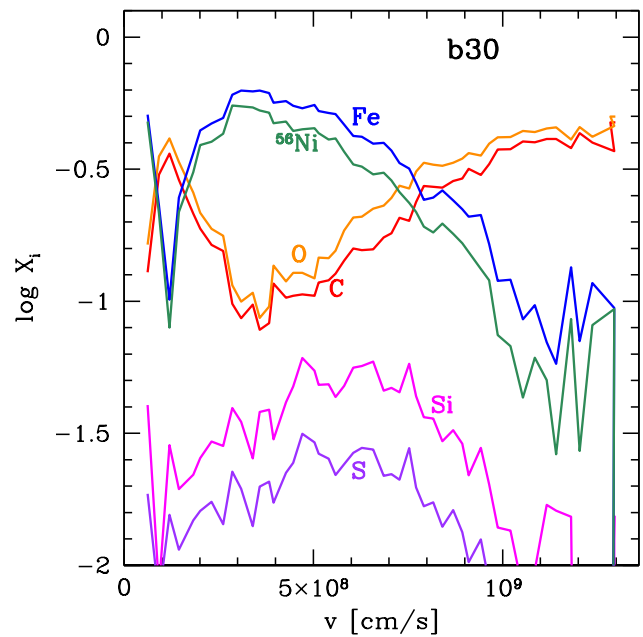
Mass fractions of most abundant elements for the model \emph{b30}
as a function of the material velocity.

\FIG rhoc3 rhoc3eve90 []
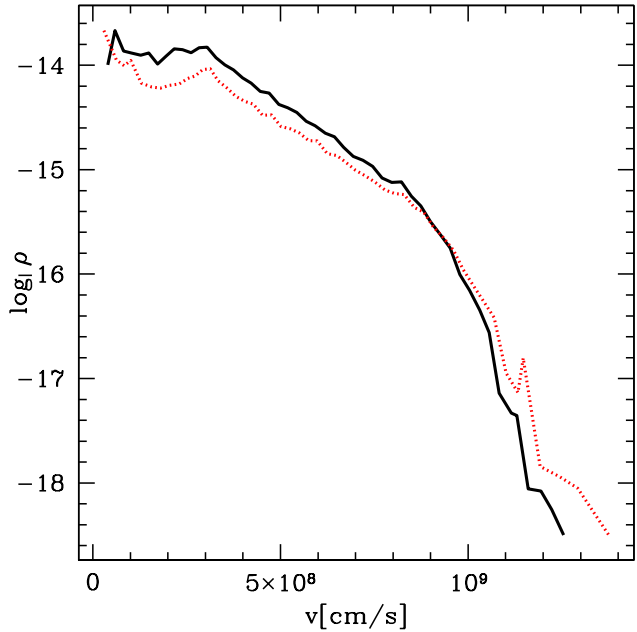
Initial ($t=10^4$ sec, black solid) and final ($t=90$ days, dotted) 
density profiles as a function of material velocity scaled to the 
same maximum for \emph{c3\_3d}.

\section{Predicted observables and their diversity}
\label{observ}

Observations of SNe~Ia indicate that a valid explosion model
should release $\sim$$10^{51}$ erg of energy and 
synthesize $\sim0.4\ldots0.7\, M_{\sun}$ of $^{56}$Ni. 
However, there exists a large diversity in the
observations ranging from the class of sub-luminous SNe~Ia 
(SN~1991bg) to the bright events (SN~1991T). The deflagration
models use in this study release  
$4.4 \ldots 6.3 \times 10^{50}\,\mathrm{erg}$ of asymptotic kinetic energy
into the ejecta and produce $0.24 \ldots  0.42 \, M_{\sun}$ of \nifsx
(see Table~\ref{energ_tab}). Thus they fall into the lower range of observational
expectations, but as of yet, do not account for the more luminous events.

The synthetic light curves are sensitive to the energy release, the
$^{56}$Ni production and the distribution of elements in the ejecta. 
Fig.~\ref{mb30c3lc} displays the computed $B$- and $V$-band light curves 
for all four of the explosion models (see Table~\ref{models_tab}). 
During maximum light these two passbands carry the main contribution 
to the luminosity. 
It is clear that in spite of modest values of explosion energy and 
$^{56}$Ni mass, these deflagration models quite nicely reproduce the 
observed absolute peak $B$- and $V$-band magnitudes and decline rates 
of observed weak and normal SN~Ia. Below we compare the computed light
curves with some well-observed nearby SNe~Ia.     

\subsection{Broad-band photometry}

Detailed comparisons with observations of SN~1992A \citep{suntzeff96}
and SN~1994D \citep{richmond95,meikle96,smith00}
are presented in Figs.~\ref{mags133_94D}-\ref{magsb30_92A}.  
These two SN~Ia were selected  
because they are both well observed and the \nifsx masses derived
for each of them \citep{stritzinger05} are similar to the values of 
the models.\footnote{The light curve templates of SN~1992A have less 
coverage in the $R$- and $I$-bands because there are fewer premaximum 
photometric points compared to in the optical passbands.} 
The observational template light curves are obtained as described in
\citet{Contardo2000} and \citet{StrLeib05}.
The filter transmission functions used to derive the light curves from the
modeled spectrum are from \citet{Stritz05}.
The zero time on the plots is for the time of $B$-band maximum 
(both for theory and observations). 

In Fig.~\ref{mags133_94D} we compare synthetic light curves
derived from model \emph{1\_3\_3} (\citealt{roepke2005}) with
observations of SN~1994D.
The distance  to  SN~1994D is still controversial.
We have assumed the distance modulus 30.4 from (\citealt{DrenkRicht99})
and the total reddening assumed is  E(B-V)$_{\rm tot}$$=$$0.022$ 

The model produced only $0.24 \, M_\odot$ of $^{56}$Ni, that is why
the modeled $UBV$-band maxima are low compared with SN~1994D. 
The decline in $B$-band light curve for 15 days 
after the maximum light is also slow for a low luminosity SN~Ia.
The near infrared maxima are of the correct order, although they
lack a convincing secondary maxima.

In Fig.~\ref{mags222_92A} we compare computed light curves
derived from model \emph{2\_2\_2} (\citealt{roepke2005}) with
observations of SN~1992A. 
The distance modulus is taken to be 31.35
and the observed light curves are de-reddened assuming 
E(B-V)$_{\rm tot}$$=$$0.017$. 
\citet{Richtler2000} have shown that  
a number of different distance determinations of the Fornax cluster of galaxies (where SN~1992A is located) agree 
well with a distance modulus of 31.35$\pm$0.04 mag (18.6$\pm$0.3Mpc).
\citet{Madore98} have published a Cepheid distance to the cluster giving a distance
modulus of 31.35 $\pm$ 0.07.

The model produced $0.3 \, M_\odot$ of $^{56}$Ni. This is not
a sufficient amount to account for the peak $B$- and $V$-band magnitudes.
However, the peak value in the $U$-band is almost reached,
although somewhat earlier than what is required by the observations.
The decline in the $B$- and $V$-bands and the shapes of $I$- and $R$-band
light curves are similar to those for model \emph{1\_3\_3}.

Fig.~\ref{magsc3_94D} displays the theoretical light curves for model 
\emph{c3\_3d} as compared with observations of SN~1994D.
The \nifsx mass is rather low ($0.28 \, M_\odot$) and the general behavior
of the fluxes is similar to the previous cases.

The best agreement with observations in the $B$- and $V$-bands is achieved with
model \emph{b30}. Fig.~\ref{magsb30_92A} shows the 
synthetic light curves derived from model \emph{b30} with observations
of SN~1992A.
Near maximum light there is a very good (almost perfect) agreement in the 
$B$- and $V$-bands as well as in the decline rate 20 days past maximum.
This is an important result for cosmological applications, since the
decline rate in those passbands (which are the main contributors 
to luminosity at this epoch) is used for calibration of the absolute
fluxes.
The flux in the $U$-band peaks early while at the same time
overshoots the observed peak luminosity, and then declines too fast.
This behavior may be explained by the significant amount of 
mixing of \Nif~in model \emph{b30} (see Fig.~5).  
With a large fraction of mixed \Nif, one would expect an excess of 
$U$-band flux prior to maximum followed by a rapid evolution. 
Less mixing of \Nif~may help to put the model in a better agreement with 
observations.
Note, however, the maximum $U$-band luminosity is on the correct order for
SN~1994D. 

Our experiments with other models show that a
more layered structure of chemical composition
may also help to produce secondary maxima in the $R-$ and $I$-bands,
which are not reproduced by the strongly mixed model in 
Figs.~\ref{mags133_94D}-\ref{magsb30_92A}. 
See Fig.~3. in \citet{blsorprt} for the abundance profile of the 
classical W7 model (\citealt{nomotow7}).

\FIG mb30c3lc mb30c3lc []
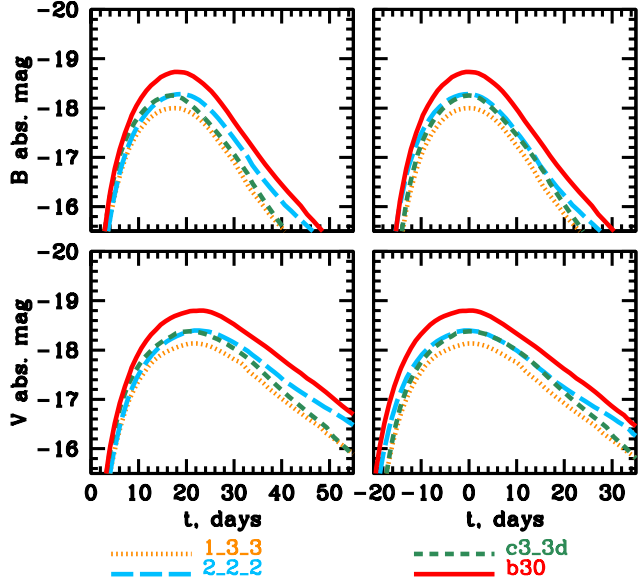
$B$- and $V$-band light curves for all 4 models.
{\em Left:} zero time is the moment of explosion.
{\em Right:} zero time is the maximum light in the given band.

\FIG mags133_94D magslc133mpa94D []
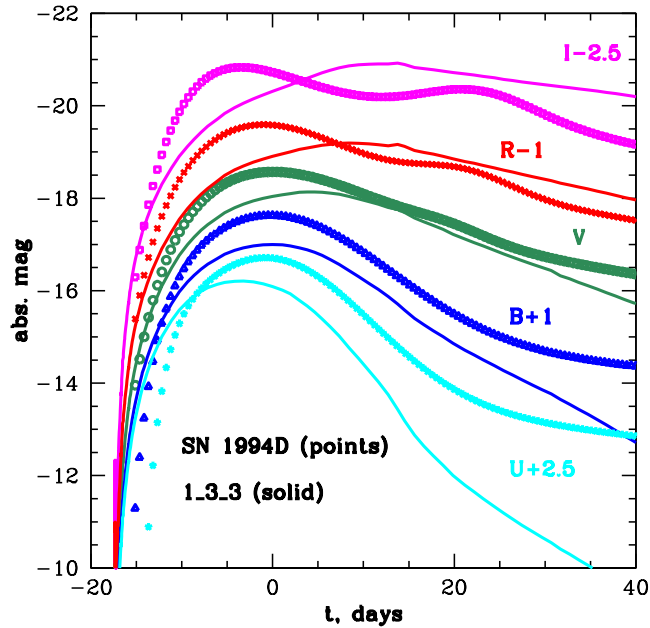
$UBVRI$ light curves for \emph{1\_3\_3} compared
  with  observations of SN1994D.
Zero time is $B$-band maximum.

\FIG mags222_92A magslc222mpa92A []
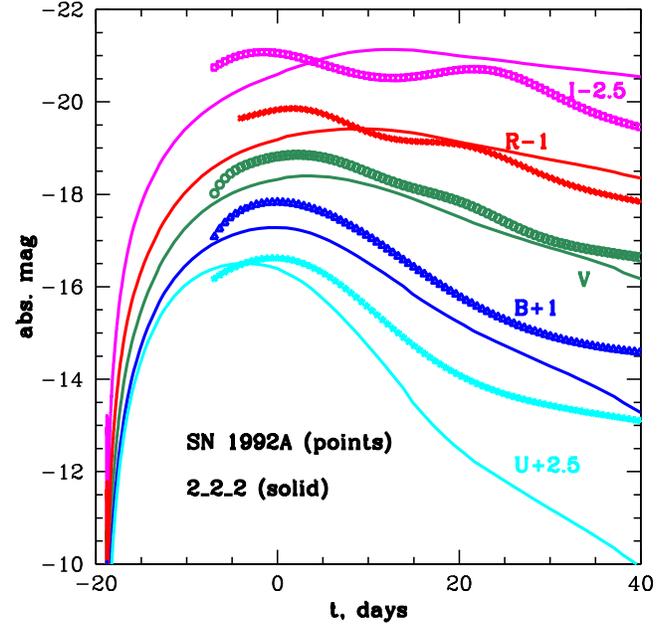
$UBVRI$ light curves for \emph{2\_2\_2} compared
  with observed light curves from SN~1992A (data points).
Zero time is $B$-band maximum.

\FIG magsc3_94D magsc3mpa94D []
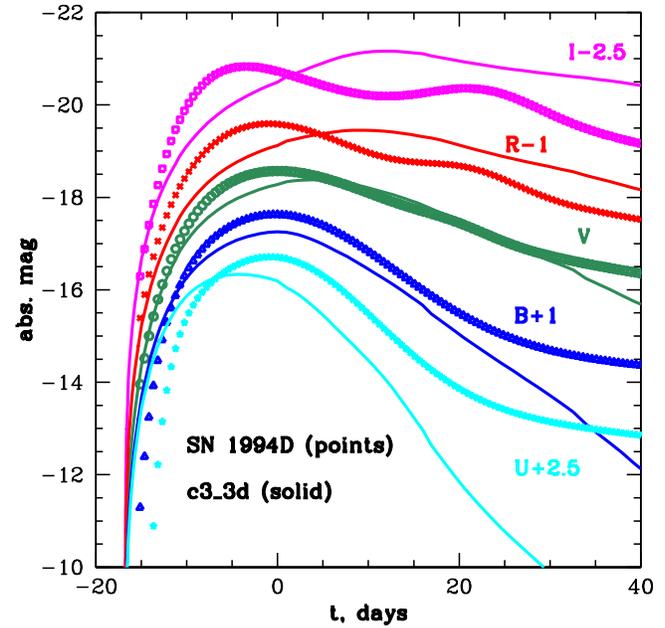
$UBVRI$ light curves for \emph{c3\_3d} and observations of SN1994D.
Zero time is $B$-band maximum.

\FIG magsb30_92A magsb30mpa92A []
Synthetic light curves derived from model \emph{b30} (solid curves)
(\citealt{Trav2004}) compared
 with observed light curves from SN~1992A (data points).
Zero time is with respect to $B$-band maximum.

Some computed light curves parameters are summarized in Table~\ref{Lc_tab}.
If we compare our results for $\Delta$m$_{15}({\rm B})$ with the Phillips 
relation 
(\citealt{phillips99}),
we find that model \emph{b30} fits quite well, while
other lower \nifsx mass models are somewhat under-luminous for their
values of $\Delta$m$_{15}({\rm B})$. In this comparison, one should take into account that we have used 
H$_{\circ}=72$ km s$^{-1}$ Mpc$^{-1}$, while \citet{phillips99}
used H$_{\circ}=65$  km s$^{-1}$ Mpc$^{-1}$.

\begin{table}
\centering
\caption{Model $B$-band light curve rise-times and $\Delta$m$_{15}({\rm B})$
\label{Lc_tab}}
\setlength{\extrarowheight}{2pt}
\begin{tabular}{llll}
\hline\hline  
model & $t_{\max}(B)$, days &    $B_{\max}$ &   $\Delta$m$_{15}(B)$   \\
\hline
\emph{1\_3\_3}  &  $17.58$ & $-18.00$ & $1.52$\\
\emph{2\_2\_2}  &  $18.59$ & $-18.28$ & $1.40$\\
\emph{c3\_3d}   &  $16.96$ & $-18.26$ & $1.46$\\
\emph{b30}      &  $18.18$ & $-18.73$ & $1.43$\\
\hline
\end{tabular}
\end{table}

\subsection{Bolometric light curves}

In Fig.~\ref{bolb30} we compare the computed bolometric light curve for the 
model \emph{b30} with several of the UltraViolet Optical InfRared ({\em UVOIR})
light curves presented in \citet{stritzinger05}.
Note that the theoretical bolometric light curve (dotted line)
is computed in the wavelength range from extreme ultraviolet to far infrared. 
The templates UVOIR light curves are computed in the
range from $U$-band (3100~\AA) to the near infrared (10,000~\AA).
Therefore if we take a simple cut from the multi-group spectrum and sum
the flux we obtain a UVOIR light curve (solid line in Fig.~\ref{bolb30})
that can be directly compared to the template UVOIR light curves.
Although the luminosity at maximum and the rise-time are in good
agreement with SN~1992A the decline
rate is somewhat slower than observed. This is mostly due to
the slow decline of the near infrared flux in the theoretical radiative
model.

In Fig.~\ref{bolc3} we compare the bolometric light derived from
model \emph{c3\_3d} to the template UVOIR light curves. 
Compared to the previous model,
\emph{c3\_3d} has a lower explosion energy and produces a 
smaller amount of \nifsx (see Table~2). As a results we see that
the total luminosity is less than what is determined from model \emph{b30}.

In the next section we compare the UVOIR luminosity for our four 
models to the luminosities derived from the template light curves by 
the method presented in \citet{StrLeib05}.
 
\FIG bolb30 Lmstrizb30 []
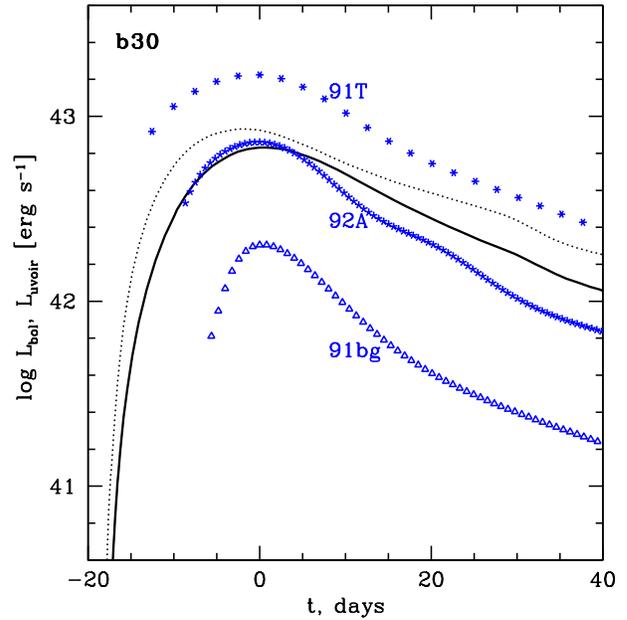
Total bolometric luminosity (dotted) and UVOIR luminosity (solid) for the model \emph{b30}.
Observations, shown for three representative SNe~Ia, should be compared
only with the solid line. Zero time is the epoch of maximum luminosity for 
each light curve.

\FIG bolc3 Lmstrizc3 []
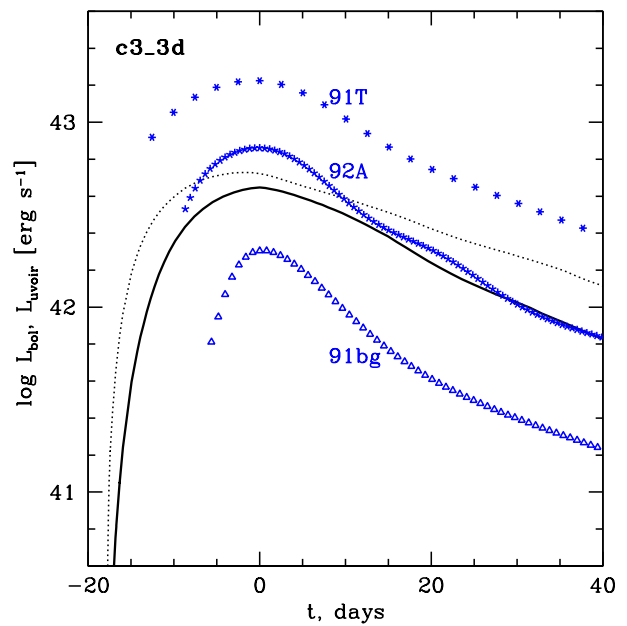
Bolometric luminosity (dotted) and UVOIR luminosity (solid) for the model 
\emph{c3\_3d.}

\FIG dep depb30 [] 
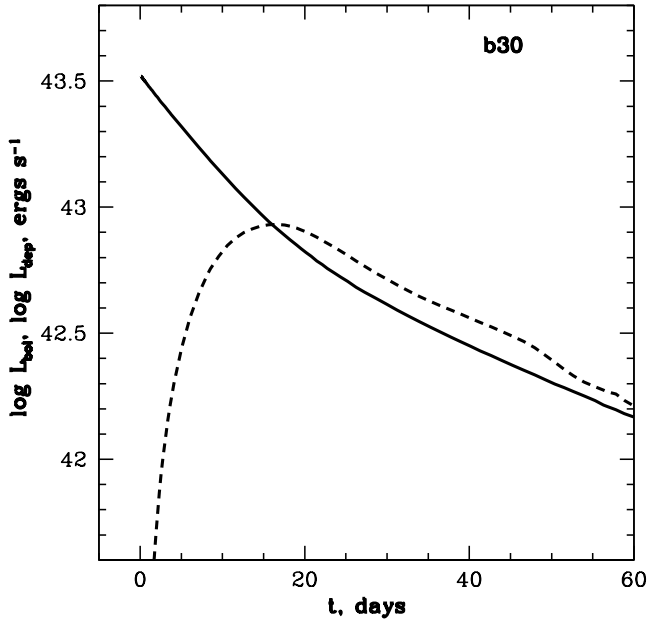
 Deposited gamma-ray (solid line) and bolometric luminosity (dashes)
 for the model \emph{b30}. Zero time is the explosion epoch.

\subsection{Photospheric velocity}

Our radiation hydro models allow us to evaluate the velocity of
material at the level of the photosphere.
We define this level as a layer with an optical depth of $2/3$ in the
$BV$-bands.
The results for the velocity at this level are shown in Fig.~\ref{vphot}.
We find that the photospheric velocity is near $\sim 10^4$ km s$^{-1}$ and
changes weakly
in the models, while observations sometimes show larger velocities before
maximum followed by a faster evolution.
For example, several days before $B$-band maximum the Si~II $\lambda6355$ absorption
feature in SN~1992A indicated velocities on the order of 
$13\times 10^3$ km s$^{-1}$ (\citealt{kirshner93}).
Moreover \citet{patat} obtain for SN~1994D $14\times 10^3$ km~s$^{-1}$ 
10 days before $B$-band maximum.
Similar values are found by \citet{branch2005}, see Fig.~\ref{vphot}.
In both SNe~Ia the velocity fell quickly near maximum light and stayed
$\sim$10$^{4}$ km~s$^{-1}$ around day 20 after the maximum.
Note, that in Fig.~\ref{vphot} the time is given from
the explosion epoch, and epochs of $B$-band maximum are presented in 
Table~\ref{Lc_tab}.

\FIG vphot  vph3models94D [] 
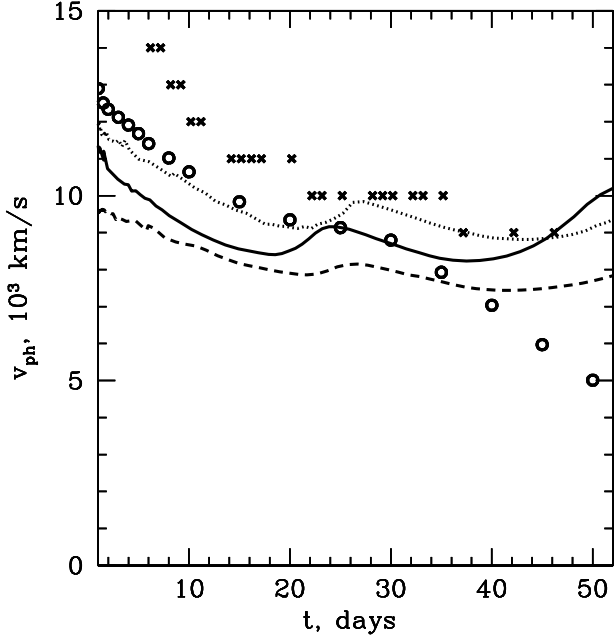
Photospheric velocity for \emph{c3\_3d} (solid), \emph{2\_2\_2} (dashed)
and \emph{b30} (dotted) defined as $R_{\tau(BV)=2/3}/t$. Circles show
the  velocity at `Rosseland' photosphere for \emph{b30}.
Crosses are the data for SN~1994D from  \citealt{branch2005}.
Zero time is the explosion epoch.

It should be noted that \citet{blondin} recently presented a detailed analysis of 
line profiles for a large sample of  SNe~Ia.
Their study indicates that there are many nuances at play that
make the procedure 
of using line profiles as a measure of photospheric velocities a difficult
task.
In addition some SNe~Ia exhibit a very
slow evolution of velocity measured by Si~II $\lambda6355$ absorption
near $\sim 10^4$ km s$^{-1}$, see Fig.~6 in \citep{blondin}.

This may be interpreted as the signature of lower
bound of Si distribution \citep[e.g., ][]{branch2005}, not as the signature of
the slow evolution of the photospheric velocity.
To illustrate this we plot in Fig.~\ref{vphot} the behaviour of the velocity on the photosphere determined by the Rosseland mean opacity for \emph{b30}. While the the ejecta
cool down after the maximum light the Rosseland opacity is determined by
red wavelengths where the matter is more transparent than in blue.
Here the photospheric velocity goes down monotonically which is usually
observed in spectroscopic studies.
The non-monotonic behaviour of the  photospheric velocity in Fig.~\ref{vphot} may be explained by different level of excitation caused by the presence of Ni56 in outer layers. This may be enough to support optical depth in bluer wavelengths,
but not sufficient for a `true' photosphere.

The observation of fast features in the spectra of some SNe~Ia
demonstrates the need for further development of the deflagration models.
Our procedure of averaging
of the original 3-D models over solid angles somewhat suppresses the fastest motions.
As a result the outermost mesh zone in model \emph{b30}
contains a mass of $8.5\times 10^{-4} M_{\sun}$ with an asymptotic velocity 
$12.98 \times 10^3$ km s$^{-1}$, while some of the tracer particles of the
3-D model had $13.81 \times 10^3$ km s$^{-1}$.

\section{Deriving $^{56}$Ni masses from SNe~Ia light curves}
\label{ni56}

Following ``Arnett's rule'' (\citealt{arnett1979,arnett1982})
one can derive the  $^{56}$Ni mass from the peak luminosity of a SN Ia. 
Arnett's rule simply states 
that at the epoch of maximum light the peak luminosity is equal to the 
rate of gamma-ray deposition inside the ejecta.
The derivation of this statement is based on many simplifying assumptions,
yet it is well satisfied in the current set of models (see Fig.~\ref{dep}).

To derive a \nifsx mass from the peak luminosity one also has to know 
the fraction of the total radioactive power that is actually deposited,
and the fraction that escapes. 
An empirical procedure for finding the \nifsx mass has been developed by 
\citet{StrLeib05}.
With a UVOIR light curve and the simple relation 
$L_{\rm max}= 2\times 10^{43} M_{\rm Ni}/M{\sun}$
erg s$^{-1}$, (and 10\% correction taking into account the difference
of UVOIR and bolometric luminosity)
they were able to make estimates of the \Nif~mass for 
a large number of SNe~Ia. 

The accuracy of this procedure may be tested with our 
synthetic light curves.
Fig.~\ref{mpabolb30} contains the comparison of the UVOIR light curve 
derived from theoretical $UBVRI$ fluxes applying the
procedure of \citeauthor{StrLeib05} with the results for model \emph{b30}. 

\FIG mpabolb30 mpaLbolb30 []
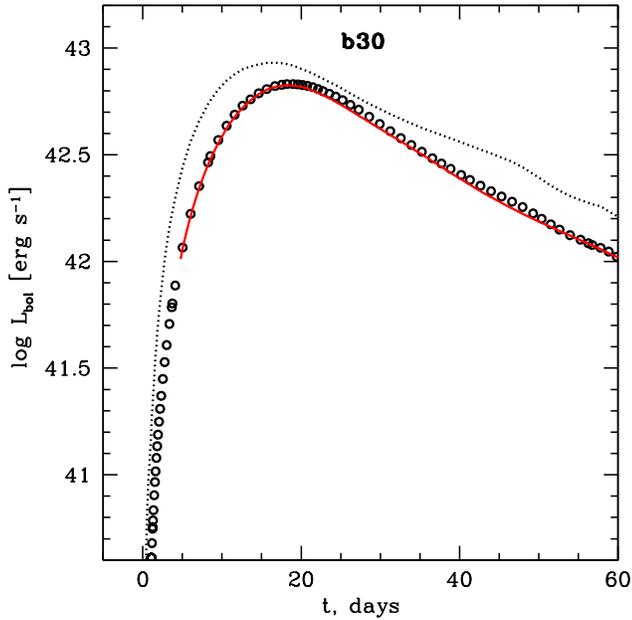
Computed UVOIR light curve for the model \emph{b30} (circles) 
compared with the UVOIR light curve derived using
the procedure of \citet{StrLeib05} (solid line) with the modeled $UBVRI$ 
light curves. The full bolometric luminosity is given by the dotted line. 
Zero time is the explosion epoch.

Table~\ref{ni56_tab} shows that the derived mass of \nifsx is
systematically lower than the actual values in the models.
Fig.~\ref{LumNi} explains the reason for this: the relation 
$L_{\rm max}= 2\times 10^{43} M_{\rm Ni}/M{\sun}$
erg s$^{-1}$ is good for true (total) bolometric luminosity
which is systematically higher than the UVOIR luminosity and the 10\%
correction is a bit small to fully account for that.

The model  \emph{b30} produced $0.42$ M$_{\sun}$ of $^{56}$Ni and a
comparable value of $0.40$ M$_{\sun}$ of $^{56}$Ni was found for SN~1992A 
by Stritzinger (2005).
The agreement is better, than suggested by numbers in the Table~\ref{ni56_tab},
because the observed UVOIR luminosity at maximum light is somewhat higher than 
in the model \emph{b30}, since our models have
IR fluxes too low at UVOIR maximum light 
(see Figs.~\ref{magsb30_92A}and~\ref{bolb30}).

\FIG LumNi LumNi []
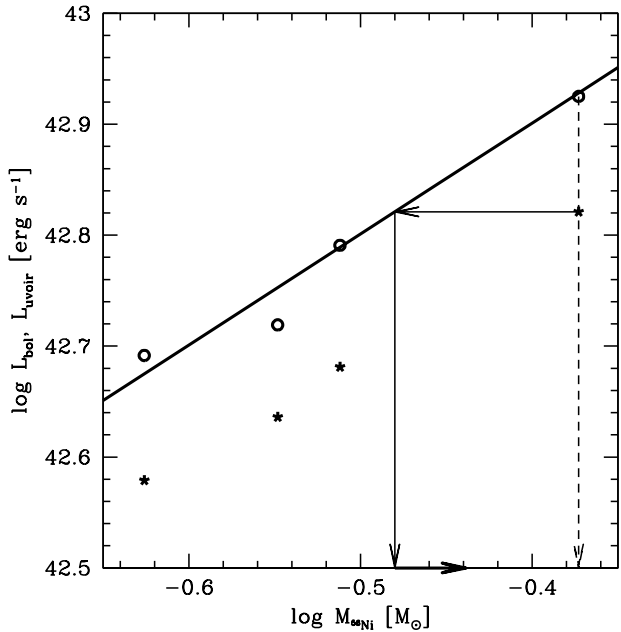
Peak values of UVOIR (stars) and bolometric (circles) luminosity
versus \nifsx mass in logarithmic scale for our four models.
Solid line is the relation from \citep{StrLeib05}.
Dashed arrow points to the correct value of \nifsx mass for 
\emph{b30}.
Thin solid line with arrow points to the lower \nifsx mass found
from UVOIR luminosity and thick solid arrow is 10\% correction
used \citep[according to][]{StrLeib05} to obtain the number given in the Table~\ref{ni56_tab}.

\begin{table}
\centering
\caption{Values of \nifsx mass derived from the theoretical 
       light curves by the method from \citet{StrLeib05} .
\label{ni56_tab}}
\setlength{\extrarowheight}{2pt}
\begin{tabular}{llll}
\hline\hline
model & $\log L_{\max}$, erg s$^{-1}$ & 
   $M(^{56}\textrm{Ni}) [M_{\sun}]$,&
 $ M(^{56}\textrm{Ni}) [M_{\sun}]$,   \\
      &  & actual & derived \\
\hline
\emph{1\_3\_3}  & $42.579 $  & $0.24$ & $0.21$\\
\emph{2\_2\_2}  &  $42.681$ & $0.31$ & $0.26$\\
\emph{c3\_3d}   &  $42.636$ & $0.28$ & $0.23$\\
\emph{b30} &  $ 42.821 $ & $0.42$ & $0.37$\\
\hline
\end{tabular}
\end{table}

\section{Discussion of the results}
\label{discuss}

This study does not produce 
theoretical light curves that perfectly match all the observations.
However, the synthetic modeled $UBV$ light curves do show 
good agreement with the observed ones for modestly
luminous SNe~Ia, although the evolution of the luminosity is somewhat
slower (particularly in the  bolometric light curve) than observed.
In this section we discuss the approximations made in the current study
and their probable influence on the results.

\subsection{Shortcomings of the current modeling}

Partially, the remaining differences between synthetic and
observed light curves can be attributed to the the explosion
models. All centrally ignited models possess an 
artificial flame ignition configuration that is known to give rise
to weak explosions. In addition in the four explosion models 
used, thermonuclear
burning was not followed below densities of 10$^{7}$ g~cm$^{-3}$.
In a more realistic explosion model burning should
continue to far lower densities. This has two effects that are likely
to improve the agreement between the modeled and observed light curves.
First, at lower densities more intermediate mass elements (IMEs)
(e.g. Si, S, and Ca) will be produced which are clearly under-abundant
in the current set of models. 
Second, burning to lower densities releases more 
nuclear energy resulting in higher expansion velocities.
Currently work is in
progress to analyze the effects of extending the burning to lower fuel
densities \citep[see][]{roepke2005a}.

It is clear that  the radiative transfer models 
need considerable improvement at certain points.
While up to 20 days after the maximum light we find good agreement 
of the best model with observations in the $B$- and $V$- bands, 
it is clear that at later epochs large deviations are present.
During these epochs when the SN~Ia ejecta enter the nebular phase,
the LTE assumption is clearly not valid. 
In addition, the procedure of averaging a 3-D model for 1-D transport 
can introduce an error which can only be corrected in a full 3-D radiative 
transfer simulations.

Another factor that has a dramatic influence on the modeled light curves  
is the treatment of the opacity.
Direct numerical summation of individual lines, as done in atmospheric codes like
{\sc phoenix} (\citealt{BHNB}) may be dangerous 
in deep interiors of SN ejecta, since the difference of the source
function and mean intensity is very small where the lines are strong.
The subtraction of the nearly equal numbers always leads to loss of accuracy,
and the result significantly depends on the order of summation of lines:
one has to start from the weakest ones  (cf. \citealt{SBring02}). 
Opacity sampling (OS) or opacity distribution function (ODF) and
$k$-distribution procedures used in static atmospheres must be modified for applications in supernova envelopes (\citealt{Wehrse02}).
However, the recipes for the statistical treatment of lines were developed by  
\citet{Wehrse02} only for the flux equation, and at present it is unclear
how to generalize them for treatment with the energy equation, e.g. Eq.~(\ref{tempr}).
A deterministic approach for an average expansion opacity to be used in the 
energy equation  was previously developed
(\citealt{SBring02}), but it is more computationally demanding compared to the 
simple formulae (\citealt{EP93}) employed in this study.

For some runs in the current set of models we have applied the new
recipes by \citet{SBring02} and found for them a small difference in comparison
with the prescription by \citet{EP93}. 
This is probably due to strong mixing of the models.
More layered models produce clear secondary maxima in the $R$- and $I$- bands, and
the use of `new' opacities from \citet{SBring02} makes them more pronounced.

Recent papers by \citet{branch2005,stehle05}
used spectroscopic analysis in different approximations and found that the composition structure in SNe~Ia is stratified but has some degree of mixing.
The latter is not so strong as the mixing in the current models.
This is another hint for need of producing more chemical stratification in the
hydrodynamic models. However, one should first build detailed NLTE spectroscopic
models for the current models. The line spectra are sensitive not only 
to the distribution of chemical elements but also to the conditions
of excitation which may change in radius in more pronounced way than the
chemical composition.

The absolute magnitudes for SNe~Ia in the infrared $J,H$ and $K$-bands is very 
uniform even for peculiar events (\citealt{krisciunas01,krisciunas04}).
This is a rather puzzling result given the variety of morphologies displayed in
the $R$- and  $I$- bands.
The opacity physics in our radiation hydro model is presently not complete in the 
$J,H$ and $K$-bands and it will be necessary to address this problem in future work.

\subsection{Hydrodynamic effects in SN~Ia}
\label{hydroia}

Progenitors of SNe~Ia are believed to be compact degenerate stars.
The explosion develops on a time-scale of a second, and homologous expansion
should be a good approximation after a few seconds.
This approximation is usually exploited in the radiative transfer
codes that neglect hydrodynamics (\citealt{EP93,Lucy05}).
However, 
\citet{PE00a} pointed out that 
the energy released in the \nifsx decay can influence the
dynamics of the expansion. The \nifsx decay energy is
$3\times10^{16}$\ergg which is equivalent to a speed of $2.5\times 10^3$ 
km s$^{-1}$, if transformed  to the kinetic energy of a
gram of material. \citet{PE00a} state
``Since the observed expansion velocity of SNe~Ia is in 
excess of $10^4$ \kms, we expect that this additional source
of energy will have a modest, but perhaps not completely negligible,
effect upon the velocity structure.''

In reality the heat released by the \nifsx decay will not all go 
into the expansion of the SN~Ia. 
If the majority of \nifsx is located in the central regions of the ejecta then
the main effect is an increase in the entropy and local pressure
(both quantities are dominated by photons in the ejecta for the first several week).
The weak overpressure will lead to a small decrease in density at the 
location of the `nickel bubble', as well as to some acceleration
of matter outside the bubble.
It is often said that the expansion of the ejecta is supersonic and pressure
cannot change the velocity of the matter, but one should remember that
in the vicinity of each material point we have a `Hubble' flow so differential
velocities are in fact {\em subsonic} in a finite volume around each point.

In the case of a more uniform distribution of \nifsx, like in the models under
consideration, we should not expect the formation of a nickel bubble,
but the general effect upon the velocity should be anticipated.

In order to compute the hydrodynamics correctly to find small
deviations from homologous expansion we should not assume that
energy conservation can be simplified to
the condition of thermal equilibrium, and
that the gas terms $dU + pdV$ can be neglected.
This simplification, when the term $4\pi \int_0^\infty
 (\eta_\nu-\alpha_{\nu}J_\nu)d\nu$  in the
co-moving frame is equal to the rate of radioactive energy deposition $\varepsilon$ is always done in models that assume homologous expansion (\citealt{arnett1980,PE00a,Lucy05}).

We do not make simplifying
approximations such as neglecting the gas terms $dU + pdV$, nor do we
assume homologous expansion.
Eq.~(\ref{tempr}) automatically accounts for any deviations from 
thermal equilibrium.

The effect is modest, but as expected it may result in a $\sim$$10$\% difference
in velocity. This effect is evident for the \emph{c3\_3d} model (see Fig.~\ref{rhoc3}). 
The solid black line is the
initial model scaled to our result at 90 days 
since explosion. We see that the density
profile has changed due to the $^{56}$Ni and \cofsx decays; it clearly 
deviates from homology. 
The growth of the velocity in outer layers is also visible.
The effect upon the total kinetic energy is smaller (see Table~\ref{energ_tab}).

All of the models considered here have strongly mixed $^{56}$Ni.
Other models, with modest mixing do not show appreciable increase of the velocity
in outer layers, but they demonstrate that
the nickel bubble, i.e. the depression of density in $^{56}$Ni-rich layers, continues to 
grow during the coasting stage. 
The change of the density is important for the deposition of gamma-ray energy,
and this is reflected in the light curve.

\subsection{Secondary maxima in $R$, $I$.}

Most SNe~Ia (except for subluminous SN~1991bg-like ones) display pronounced secondary maxima 
in their near-infrared and infrared light curves. 
Observations show a large variety of observed infrared maxima.
According to \citet{nobili05}, the light curve in $I$-band can peak 
before, as well as after $B_{\rm max}$ (between $-3$
days and $+3$ days with respect to $B_{\rm max}$). On average, the secondary maxima
occurs 23.6 after the $B_{\rm max}$ with a dispersion of $\sigma=4.4$ days.
However, there are strong arguments that this behavior is not random.
\citet{nobili05} found a correlation between the time of the secondary peak 
and the $B$-band stretch parameter. A luminosity-width relation was also
found between the peak $I$-band magnitude and the $B$- and $I$-band
stretch parameter. \citet{Contardo2000} found some similar 
correlations between the strength of the secondary maximum and times 
of derivatives of the light curve.

This is a very interesting and long-standing problem for the theory of
light curves.
Although \citet{EThn} previously demonstrated that some models
do produce a secondary infrared maxima (in that case it was
a sub-Chandrasekhar model), the theory is still far from providing robust
predictions concerning the shape of $I$-band light curve and the 
correlations pointed out in \citet{nobili05}.

For example, \citet{hoeflich95} found a weak secondary maxima for W7, 
and rather noisy $I$-band light curves for other models.
\citet{PE00b} also found rather weak secondary maxima, depending strongly on 
the atomic line database, nevertheless \citet{EThn} and \citet{PE00b} 
have presented a physical
argument that attempts to explain the observed secondary maxima.
According to them, after $\sim$ 20 days past maximum light, when the monochromatic opacity 
in the near infrared becomes small due to recombination of higher ions, there exists
a large amounts of singly ionized emitters (e.g. Ca~II, Fe~II, and Co~II).
Through fluorescence these lines cause an enhanced 
`leakage' of previously
trapped photons, thus giving rise to the secondary maxima. 

For this mechanism to produce a secondary maxima 
the SN~Ia model must require a large mass fraction of
Ca, Fe, and Co located in central layers that are emitting photons 
near 20 days past maximum light. 
Too strong mixing of iron-peak elements reduces their amount
in the center. With the MPA models the total mass of Ca 
is relatively small. Model \emph{b30} has 0.0037 M$_{\sun}$ of Ca, while
DD4 contains 0.0375 M$_{\sun}$ (\citealt{PE00b}) and W7 contains
0.025 M$_{\sun}$; the latter two containing an order of magnitude more!

As previously mentioned, less pronounced mixing may allow 
model \emph{b30} to produce $R$- and $I$-band light curves that agree with the
observation more closely. With less mixing of Fe-group elements and IMEs 
like Ca, our models may produce $R$- and $I$- band secondary maxima.
This may also lead to a delay in the $U$-band rise-time. 

What should be changed in the deflagration models to
decrease the degree of the mixing, but at the same time to increase the
amount of \nifsx?
We believe that the only feasible way is to lower the ignition density considerably. This will have two effects. First, electron captures become
unimportant, increasing the \nifsx at the expense of the other
NSE nuclei. Secondly, at low densities the $^{12}$C-$^{12}$C burning rate is low,
too, and the heat diffusion time gets longer, thus reducing
the laminar burning speed and giving the star more time to
expand before rapid turbulent burning sets in. This should in fact
also reduce the amount of mixing. These conjectures have to be tested
in future detailed numerical modeling.

\section{Conclusions}
\label{concl}

How the diversity of local SNe~Ia can affect their use as distance indicators and
the possibility of systematic trends (as a function of redshift) in their observed 
properties cannot be addressed without detailed radiative transfer modeling. 

The set of hydrodynamical models used in this studied synthesized modest 
amounts of \nifsx $\sim$0.3 - 0.4 M$_{\sun}$ and displayed relatively low
explosion energies (lower than the `standard' 1.5 foe expected for a `typical' SN~Ia).
However, our results show that Chandrasekhar mass models that burn by 
a pure deflagration do produce both $UBV$ light curves and photospheric expansion 
velocities that match well with observed {\em weak to normal} SNe~Ia.
As the majority of flux during maximum light is emitted within these passbands 
this is an encouraging result. Moreover, it is these passbands that are of great
importance for cosmological applications of SNe~Ia.
It is clear that the models require some improvement in explaining the shapes of
the near infrared light curves and in explaining fast spectral features that are
observed in many normal events. 
The bolometric light curves calculated from our deflagration models 
evolve slightly slower than what is indicated from observations. 
These discrepancies hint at the necessity of producing
faster moving ejecta, and somewhat less mixed chemical composition.
The latest deflagration (to be discussed in a subsequent publication)
models are promising in this aspect.

From the point of view of radiative transfer modeling several issues must be dealt with
including: (1) determining the sensitivity of the results to the completeness of 
the atomic line data base, (2) the robustness of the various approximations employed
such as the expansion opacity prescriptions, and (3) determining the importance of 
(time-dependent!) NLTE effects. These issues are currently being addressed and small
steps are being made to bring us closer to a full 3-D transport model for SNe~Ia.

\begin{acknowledgements} 
This work was supported in part by the European Research
Training Network ``The Physics of Type Ia Supernova Explosions'' under
contract HPRN-CT-2002-00303.
SB and ES are supported by MPA and UCSC guest programs and
partly by the Russian
Foundation for Basic Research (projects  05-02-17480, 04-02-16793). Special thanks
to Nick Suntzeff for access to unpublished SN~1992A photometry.
We thank the referee for his thorough report. 
\end{acknowledgements}


\begin{thebibliography}{64}
\expandafter\ifx\csname natexlab\endcsname\relax\def\natexlab#1{#1}\fi

\bibitem[{{Arnett}(1979)}]{arnett1979} Arnett, W.~D.\ 1979, \apjl, 
230, L37 
 
\bibitem[{{Arnett}(1980)}]{arnett1980}
  {Arnett}, W.D. 1980, ApJ, 237, 541 

\bibitem[{{Arnett}(1982)}]{arnett1982}
{Arnett}, W.~D. 1982, \apj, 253, 785

\bibitem[{{Baron} {et~al.}(1996{\natexlab{a}}){Baron}, {Hauschildt} \& {Mezzacappa}}]{BHM} Baron, E., Hauschildt, P.~H., \& Mezzacappa, A.\ 1996{\natexlab{a}}, \mnras, 278, 763 

\bibitem[{{Baron} {et~al.}(1996{\natexlab{b}}){Baron}, et~al.}]{BHNB} Baron, E., Hauschildt, 
P.~H., Nugent, P., \& Branch, D.\ 1996{\natexlab{b}}, \mnras, 283, 297 

\bibitem[Benetti et al.(2004)]{benetti04}
Benetti, S., et al. 2004, \mnras, 348, 261
 
\bibitem[{{Blinnikov} {et~al.}(1998)}]{blinn98} Blinnikov, S.~I., 
Eastman, R., Bartunov, O.~S., Popolitov, V.~A., \& Woosley, S.~E.\ 1998, 
\apj, 496, 454 
 
\bibitem[{{Blinnikov} \& {Sorokina}(2004)}]{blsorprt} Blinnikov, S., 
\& Sorokina, E.\ 2004, \apss, 290, 13 
 
\bibitem[{{Blinnikov} \& {Sorokina}(2000)}]{blsoruv} Blinnikov, 
S.~I., \& Sorokina, E.~I.\ 2000, \aap, 356, L30 

\bibitem[{{Blondin} {et~al.}(2005) {Blondin}, et~al.}]{blondin}
  Blondin, S.,  Dessart, L.,  Leibundgut, B., et~al.\ 2005, 
astro-ph/0510089

\bibitem[{{Branch} {et~al.}(1993)}]{branch1993a} Branch, D., Fisher, A., 
\& Nugent, P.\ 1993, \aj, 106, 2383 

\bibitem[{{Branch} {et~al.}(2005)}]{branch2005} Branch, D., Baron, E., 
Hall, N.,  Melakayil, M., \&  Parrent, J.\ 2005, \pasp, 117, 545,

\bibitem[{{Chang} \& {Lingenfelter}(1993)}]{ChanLing93} Chan, K. W., \&~Lingenfelter, R. E. 1993, ApJ, 405, 614.

\bibitem[{{Colgate} {et~al.}(1980{\natexlab{a}})}]{Colgate80a} Colgate, S.~A., Petschek, A.~G., \&~Kriese, J.~T. 1980{\natexlab{a}}, in AIP 
             Conference Proceedings, No.~63:  Supernova Spectra,
             ed.~R.~Meyerott \&~G.~H.~Gillespie (New York:  American
             Institute of Physics), 7.

\bibitem[{{Colgate} {et~al.}(1980{\natexlab{b}})}]{Colgate80b} Colgate, S.~A., Petschek, A.~G., \&~Kriese, J.~T. 1980{\natexlab{b}}, ApJ, 237,
            L81.  

\bibitem[{{Contardo} {et~al.}(2000) {Contardo}, {Leibundgut} \& {Vacca}}]{Contardo2000} Contardo, G., Leibundgut, B., \& Vacca, W.~D.\ 2000, \aap, 359, 876 

\bibitem[{{Drenkhahn} \& {Richtler}(1999)}]{DrenkRicht99}
Drenkhahn, G. \& Richtler, T.\ 1999, \aap, 349, 877


\bibitem[{{Eastman}(1997)}]{EThn}
    Eastman, R.~G. In: \emph{Thermonuclear Supernovae}.
    ed. by P.~Ruis-Lapuente et al. (Kluwer Academic
    Pub., Dordrecht, 1997) p. 571

\bibitem[{{Eastman} \& {Pinto}(1993)}]{EP93} Eastman, R.~G. \& 
Pinto, P.~A.\ 1993, \apj, 412, 731 

\bibitem[{{Evans}(1998)}]{evans}  Evans,   K.~F.\ 1998, J. Atmospheric Sciences,
55, 429

\bibitem[{{Gamezo} {et~al.}(2003){Gamezo}, {Khokhlov}, {Oran}, {Chtchelkanova},
  \& {Rosenberg}}]{gamezo2003a}
{Gamezo}, V.~N., {Khokhlov}, A.~M., {Oran}, E.~S., {Chtchelkanova}, A.~Y., \&
  {Rosenberg}, R.~O. 2003, Science, 299, 77

\bibitem[{{Hillebrandt} \& {Niemeyer}(2000)}]{hillebrandt2000a}
{Hillebrandt}, W. \& {Niemeyer}, J.~C. 2000, \araa, 38, 191

\bibitem[{{Hillebrandt} {et~al.}(2000){Hillebrandt}, {Reinecke}, \&
  {Niemeyer}}]{hillebrandt2000b}
{Hillebrandt}, W., {Reinecke}, M., \& {Niemeyer}, J.~C. 2000, in Proceedings of
  the {XXXV}th rencontres de {M}oriond: {E}nergy densities in the universe, ed.
  R.~{Anzari}, Y.~{Giraud-H\'{e}raud}, \& J.~{Tr\^{a}n Thanh V\^{a}n}
  (Th\'{\^{e}} Gi\'{o}i Publishers), 187--195


\bibitem[{{H\"oflich}(1995)}]{hoeflich95} H\"oflich, P. \ 1995,  ApJ, 443, 89

\bibitem[{{H\"oflich}(2002)}]{hoeflich02} H\"oflich, P. \ 2002,
Workshop on Stellar Atmosphere
Modeling, 8-12 April 2002 Tuebingen.  Eds: I.~Hubeny, D.~Mihalas, K.~Werner,
astro-ph/0207103.

\bibitem[{{H\"oflich}  {et~al.}(1993) {H\"oflich}, {Khokhlov}, \& {M\"uller}}]{hoeflich93} H\"oflich, P., Khokhlov, A.,  \& 
M\"uller, E.   1993, A\&A, 270, 223

\bibitem[{{H\"oflich} {et~al.}(1998) {H\"oflich}, {Wheeler}, \& {Thielemann}}]{hwt98} H\"oflich, P., Wheeler, J. C.,  \& Thielemann, F. K. 1998,
ApJ, 495, 617 

\bibitem[{{H\"oflich} {et~al.}(2000){H\"oflich}, et~al.}]{hoflich00}
  H\"oflich, P., Nomoto, K., Umeda, H., Wheeler, J. C. 2000, ApJ, 528, 590

\bibitem[{{Jeffery}(1999)}]{Jeffery99} Jeffery, D.L. 1999, astro-ph/9907015.

\bibitem[{{Kirshner} {et~al.}(1993){Kirshner}, et~al.}]{kirshner93} Kirshner, R.~P., et al.\ 1993, \apj, 415, 589 

\bibitem[{{Kozma} {et~al.}(2005){Kozma}, {Fransson}, {Hillebrandt},
  {Travaglio}, {Sollerman}, {Reinecke}, {R{\"o}pke}, \&
  {Spyromilio}}]{kozma2005a}
{Kozma}, C., {Fransson}, C., {Hillebrandt}, W., {et~al.} 2005, \aap, 437, 983

\bibitem[Krisciunas et al.(2001)]{krisciunas01}Krisciunas,~K., Phillips,~M.M., Stubbs,~C. et al., 2001, \aj, 122, 1616

\bibitem[Krisciunas et al.(2004)]{krisciunas04}Krisciunas,~K., Phillips,~M.~M. \& Suntzeff,~N.~B., 2004, \apj, 602L, 81

\bibitem[{{Lucy}(2005)}]{Lucy05} Lucy, L.~B.\ 2005, \aap, 429, 19 

\bibitem[{{Madore} {et~al.}(1998) {Madore}, et~al.}]{Madore98} Madore, B.~F., et al.\ 1998, \nat, 395, 47 

\bibitem[{{Mannucci} {et~al.}(2005) {Mannucci}, {Della Valle}, \& {Panagia}}]{Mannucci}
     Mannucci, F., Della Valle, M., \&  Panagia, N. 2005,
 \mnras, astro-ph/0510315 

\bibitem[Meikle et al.(1996)]{meikle96}
Meikle, W.~P.~S., Cumming R.~J., Geballe T.~R., et~al. 1996, \mnras, 281, 263
 
\bibitem[{{Milne} {et~al.}(1997) {Milne}, et~al.}]{Milne97} Milne, P. A., The, L.-S., \& Leising, M. D. 1997, in Proc.~The Fourth Compton Symposium, ed.~C.~D.~Dermer,
             M.~S.~Strickman, \&~J.~D.~Kurfess (New York:
             American Institute of Physics Press), 1022, astro-ph/9707111.

\bibitem[{{Niemeyer} \& {Hillebrandt}(1995)}]{niemeyer1995b}
{Niemeyer}, J.~C. \& {Hillebrandt}, W. 1995, \apj, 452, 769

\bibitem[{{Nobili} {et~al.}(2005) {Nobili}, et~al.}]{nobili05} Nobili, S., et al.\ 
2005, \aap, 437, 789 

\bibitem[{{Nomoto} {et~al.}(1984) {Nomoto}, {Thielemann},\& {Yokoi}}] {nomotow7} Nomoto, K., Thielemann, F.-K., \& Yokoi, K.\ 1984, \apj, 286, 644 

\bibitem[{{Nomoto} {et~al.}(2003) {Nomoto}, et~al.}] {nomoto03}
  Nomoto, K., Uenishi, T., Kabayashi, C., Umeda, H., Ohkubo, T., Hachisu, I.
  \& Kato, M., 2003, in {\em From Twilight to Highlight: The Physics of
  Supernovae}, W. Hillebrandt \& B. Leibundgut eds., ESO/Springer series 
  ``ESO Astrophysics Symposia'' (Berlin), p.115 (astro-ph/0308138)


\bibitem[{{Patat} {et~al.}(1996) {Patat}, et~al.}]{patat} Patat, F., Benetti, 
S., Cappellaro, E., Danziger, I.~J., della Valle, M., Mazzali, P.~A., \& 
Turatto, M.\ 1996, \mnras, 278, 111 

\bibitem[Pignata et al.(2004)]{pignata04}
Pignata, G., et al. 2004, \mnras, 355, 178
  
\bibitem[{{Pinto} \& {Eastman}(2000{\natexlab{a}})}]{PE00a}
{Pinto}, P.~A. \& {Eastman}, R.~G. 2000{\natexlab{a}}, \apj, 530, 744

\bibitem[{{Pinto} \& {Eastman}(2000{\natexlab{b}})}]{PE00b} {Pinto}, P.~A. \& {Eastman}, R.~G. 2000{\natexlab{b}}, \apj, 530,  757

\bibitem[{{Phillips}(1993) }]{phillips93}
   Phillips, M.M. 1993, \apj, 413, L105

\bibitem[{{Phillips} {et~al.}(1999) {Phillips}, et~al.}]{phillips99}
 {{Phillips}, M.~M., {Lira}, P., {Suntzeff}, N.~B., {Schommer}, R.~A.  
        {Hamuy}, M. \& {Maza}, J.}, 1999,
 \aj, 118, 1766
   
\bibitem[{{Pskovskii}(1977)}]{pskovskii77}
  Pskovskii, Yu.P. 1977, Soviet Astron. 21, 675

\bibitem[{{Reinecke} {et~al.}(1999{\natexlab{a}}){Reinecke}, {Hillebrandt}, \&
  {Niemeyer}}]{reinecke1999a}
{Reinecke}, M., {Hillebrandt}, W., \& {Niemeyer}, J.~C. 1999{\natexlab{a}},
  Astron. Astrophys., 347, 739

\bibitem[{{Reinecke} {et~al.}(2002{\natexlab{a}}){Reinecke}, {Hillebrandt}, \&
  {Niemeyer}}]{reinecke2002b}
{Reinecke}, M., {Hillebrandt}, W., \& {Niemeyer}, J.~C. 2002{\natexlab{a}},
  \aap, 386, 936

\bibitem[{{Reinecke} {et~al.}(2002){Reinecke}, {Hillebrandt}, \&
  {Niemeyer}}]{reinecke2002d}
{Reinecke}, M., {Hillebrandt}, W., \& {Niemeyer}, J.~C. 2002,
  \aap, 391, 1167

\bibitem[Richmond et al.(1995)]{richmond95}
Richmond, M.~W., Treffers, R.~R., Filippenko, A.~V., et~al. 1995, \aj, 109, 2121

\bibitem[{{Richtler} {et~al.}(2000) {Richtler}, et~al.}]{Richtler2000} Richtler, T., 
Drenkhahn, G., G{\'o}mez, M., \& Seggewiss, W.\ 2000, From Extrasolar 
Planets to Cosmology: The VLT Opening Symposium, Proceedings of the ESO 
Symposium held at Antofagasta, Chile, 1-4 March 1999.~Edited by Jacqueline 
Bergeron and Alvio Renzini.~Berlin: Springer-Verlag, 259
 

\bibitem[{{Ruiz-Lapuente}(1997)}]{Pilar97} Ruiz-Lapuente, P. 1997,
             in Proc.~NATO ASI on
             Thermonuclear Supernovae, ed.~P.~Ruiz-Lapuente,
             R.~Canal, \&~J.~Isern
             (Dordrecht:  Kluwer), 681 

\bibitem[{{Ruiz-Lapuente} \& {Spruit}(1998)}]{PilarHenk98} Ruiz-Lapuente, P., \& Spruit, H. C. 1998, \apj, 500, 360

\bibitem[{{R{\"o}pke}(2005)}]{roepke2005c}
{R{\"o}pke}, F.~K. 2005, \aap, 432, 969

\bibitem[{{R{\"o}pke} \& {Hillebrandt}(2004)}]{roepke2004c}
{R{\"o}pke}, F.~K. \& {Hillebrandt}, W. 2004, \aap, 420, L1

\bibitem[{{R{\"o}pke} {et~al.}(2005){R{\"o}pke}, et~al.}]{roepke2005}
{R{\"o}pke}, F.~K., {Gieseler}, M., {Reinecke}, M. {Travaglio}, C. \& {Hillebrandt}, W. 2005,
  \aap, submitted

\bibitem[{{R{\"o}pke} \& {Hillebrandt}(2005{\natexlab{a}})}]{roepke2005a}
{R{\"o}pke}, F.~K. \& {Hillebrandt}, W. 2005{\natexlab{a}}, \aap, 429, L29

\bibitem[{{R{\"o}pke} \& {Hillebrandt}(2005{\natexlab{b}})}]{roepke2005b}
{R{\"o}pke}, F.~K. \& {Hillebrandt}, W. 2005{\natexlab{b}}, \aap, 431, 635

\bibitem[{{R{\"o}pke} {et~al.}(2005){R{\"o}pke}, et~al.}]{roepke2005d}
{R{\"o}pke}, F.~K., {Hillebrandt}, W., {Niemeyer}, J.~C. \& {Woosley},
S.~E. 2005,
  \aap, submitted

\bibitem[Smith et al.(2000)]{smith00}
Smith, C., et al. (private communication)

\bibitem[{{Sorokina} \& {Blinnikov}(2002)}]{SBring02} {Sorokina}, E.I., 
{Blinnikov}, S.I. 2002.
\emph{Nuclear
  Astrophysics, 11th Workshop at Ringberg Castle, Tegernsee, Germany,
  February 11--16, 2002}, ed. by E.~M\"uller, W.~Hillebrandt,  57

\bibitem[{{Sorokina} {et~al.}(2000){Sorokina},{Blinnikov}, \& {Bartunov}}]{Sorokina00} Sorokina, E.~I., 
Blinnikov, S.~I., \& Bartunov, O.~S.\ 2000, Astronomy Letters, 26, 67 

\bibitem[{{Sorokina} {et~al.}(2004){Sorokina}, et~al.}]{Sorokina04} Sorokina, E.~I., 
Blinnikov, S.~I., Kosenko, D.~I., \& Lundqvist, P.\ 2004, Astronomy 
Letters, 30, 737 

\bibitem[Stehle et al.(2005)]{stehle05}
Stehle, M., et al. 2005, \mnras, 360, 1231

\bibitem[Stritzinger(2005)]{stritzinger05}
Stritzinger, M. 2005, Technische Universit\"at M\"unchen Dissertation

\bibitem[{{Stritzinger} \& {Leibundgut}(2005)}]{StrLeib05} Stritzinger, 
M., \& Leibundgut, B.\ 2005, \aap, 431, 423 

\bibitem[{{Stritzinger} {et~al.}(2005){Stritzinger},et~al.}]{Stritz05}
 Stritzinger, M., 
Suntzeff, N.~B., Hamuy, M., Challis, P., Demarco, R., Germany, L., \& 
Soderberg, A.~M.\ 2005, \pasp, 117, 810 

\bibitem[Suntzeff(1996)]{suntzeff96}
Suntzeff, N.~B. 1996, in: IAU Colloquium 145, Supernovae and Supernova
Remnants, ed. R. McCray, \& Z. Wang, Cambridge University Press,
Cambridge, 41

\bibitem[{Swartz} {et~al.}(1995)]{SSH} Swartz, D.~A., 
Sutherland, P.~G., \& Harkness, R.~P.\ 1995, \apj, 446, 766 

\bibitem[{{Travaglio} {et~al.}(2004), {Travaglio}, et~al.}]{Trav2004} Travaglio, C., 
Hillebrandt, W., Reinecke, M., \& Thielemann, F.-K.\ 2004, \aap, 425, 1029 
 

\bibitem[{{Wehrse} {et~al.}(2002){Wehrse},{Baschek}, \& {von Waldenfels}}]{Wehrse02} Wehrse, R., Baschek, B., 
\& von Waldenfels, W.\ 2002, \aap, 390, 1141 
 

\end{thebibliography}
\end{document}